\documentclass[
]{ceurart}

\usepackage{pifont}
\usepackage{listings}
\usepackage[show]{chato-notes}
\usepackage{xcolor}

\newcommand{\cm}[1]{#1}
\newcommand{\sm}[1]{#1}

\def\tick{\ding{51}}
\def\cross{\ding{55}}

\lstdefinestyle{mystyle}{
    language=Python,
    morekeywords = {with, as},
    basicstyle=\ttfamily\footnotesize,
    keywordstyle=\color{blue}\bfseries,
    commentstyle=\color{gray}\itshape,
    stringstyle=\color{red},
    numberstyle=\tiny\color{gray},
    numbers=left,
    stepnumber=1,
    numbersep=5pt,
    backgroundcolor=\color{lightgray!20},
    frame=single,
    rulecolor=\color{black},
    breaklines=true,
    breakatwhitespace=true,
    captionpos=b,
    tabsize=4,
    showspaces=false,
    xleftmargin=10pt,    %
    aboveskip=5pt,      %
    belowskip=5pt,      %
    columns=flexible,    %
}
\begin{document}

\copyrightyear{2025}
\copyrightclause{Copyright for this paper by its authors.
  Use permitted under Creative Commons License Attribution 4.0
  International (CC BY 4.0).}

\conference{Second International Workshop on Open Web Search (WOWS 2025)}
  
\title{On Precomputation and Caching in Information Retrieval Experiments with Pipeline Architectures}

\author{Sean MacAvaney}[%
orcid=0000-0002-8914-2659,
email=sean.macavaney@glasgow.ac.uk,
]
\fnmark[1]
\address{University of Glasgow, United Kingdom}

\author{Craig Macdonald}[%
orcid=0000-0003-3143-279X,
email=craig.macdonald@glasgow.ac.uk,
]
\fnmark[1]

\fntext[1]{Listed alphabetically. These authors contributed equally.}

\begin{abstract}
Modern information retrieval systems often rely on multiple components executed in a pipeline. In a research setting, this can lead to substantial redundant computations (e.g., retrieving the same query multiple times for evaluating different downstream rerankers). To overcome this, researchers take cached ``result'' files as inputs, which represent the output of another pipeline. However, these result files can be brittle and can cause a disconnect between the \textit{conceptual} design of the pipeline and its \textit{logical} implementation. To overcome both the redundancy problem (when executing complete pipelines) and the disconnect problem (when relying on intermediate result files), we describe our recent efforts to improve the caching capabilities in the open-source PyTerrier IR platform. We focus on two main directions: (1) automatic implicit caching of common pipeline prefixes when comparing systems and (2) explicit caching of operations through a new extension package, pyterrier-caching. These approaches allow for the best of both worlds: pipelines can be fully expressed end-to-end, while also avoiding redundant computations between pipelines.
\end{abstract}


\maketitle

\section{Introduction}

Information retrieval systems are now often multi-stage architectures. In large-scale deployments, this allows separation of efforts: one team may work on the indexer or first-stage retriever architecture, while other teams may be responsible for rerankers (e.g. learning-to-rank, neural models), ad selection, answer generation or result presentation. While developed independently, it is imperative that these pipeline components successfully operate in tandem. For instance, if a reranker is not prepared to handle the distribution of results provided by the retriever, the engine's effectiveness will suffer.

When proposing a new method, it is important for researchers to demonstrate its robustness in different environments. Robustness is typically shown by evaluating the method on multiple benchmarks, but this is only one dimension of robustness. We argue that it is also important to show how well a method performs in various pipelines since this reflects the diversity of environments in which the method may be deployed.

However, the available data and tooling often make this type of experimentation challenging. For instance, some QA datasets (such as 2Wiki~\cite{DBLP:conf/coling/HoNSA20}) commonly provide retrieved documents for a single retriever, meaning that the impact of answer generation quality to different retrieval systems is not systematically ablated; precomputing the first-stage results for testing an answer generator or reranker can lead to an inflexible experimental workflow, that prevents the researcher from {\em dog-fooding}, i.e. testing the approach on their own queries.

A central goal of PyTerrier~\cite{DBLP:conf/ictir/MacdonaldT20,DBLP:conf/cikm/MacdonaldTMO21}\footnote{\url{http://github.com/terrier-org/pyterrier}} is to provide a shared platform that enables this kind of experimentation. Components in the platform can be composed into declarative pipelines that clearly define their constituent components, \cm{called Transformers\footnote{\sm{We note that the term \textit{tranformer} is overloaded, also referring to a neural network architecture, an electrical component, etc.}},} and how they interact. This design also allows individual components to be easily ablated (e.g., swapping out one retriever for another). However, this approach can incur substantial redundant costs at experimentation time. For instance, when comparing two answer generators under the same retrieve-and-rerank pipeline, the preceding pipeline is fully executed twice (even though the results are the same). Similarly, if the researcher seeks to ablate the retriever choice within the same pipeline, the reranker will redundantly score any document retrieved by both systems twice, even though they are assigned the same score.

This paper presents our recent efforts to address these problems while \cm{retaining} the platform's flexibility. We apply two complementary approaches. The first automatically detects common prefixes in pipelines when running an experiment and only executes the common prefix once (Section~\ref{sec:common-prefix}). The second approach is to let users explicitly define a caching strategy over individual components (Section~\ref{sec:explicit}). Together, approaches will help reduce the computational overhead of conducting experiments, abiding by the ``reuse'' principle of GreenIR~\cite{DBLP:conf/sigir/ScellsZZ22}.

Naturally, our approaches are not the only ways to perform caching for information retrieval research.\footnote{In fact, PyTerrier previously had a general-purpose caching operator ($\sim$). This operator is now deprecated in favor of these more robust and flexible caching approaches.} However, we feel that our approaches hit a sweet spot between functionality and ease-of-use for day-to-day experimentation. The traditional file-based workflow involves saving intermediate results to files (e.g., TREC-formatted ``result'' or ``run'' files). These result files can be used as starting points (and are hosted in places like ranxhub~\cite{DBLP:conf/sigir/Bassani23} or the TREC ``Past TREC Results" page\footnote{\url{https://trec.nist.gov/results.html}}), but result files on their own do not clearly define their provenance\footnote{We note that efforts have been undertaken to improve the provenance of result files, however~\cite{DBLP:conf/sigir/BreuerKS22}.} and are relatively brittle (e.g., one could mistakenly use a run file that operated over a topic's ``title'' field when the remainder of the pipeline runs over its ``description''). Meanwhile, the TIREx platform~\cite{DBLP:conf/lwa/FrobeRMDB0HP23,DBLP:conf/sigir/FrobeRMDRB0HP23} performs community-wide pipeline prefix caching, but requires containerising all components (which is a benefit in terms of reproducibility, but reduces flexibility). Finally, we note that there is a significant body of literature on the caching of search engine posting lists~\cite{baeza2007impact,10.1145/2493175.2493176} or results~\cite{gan2009improved,markatos2001caching}, however, to the best of our knowledge, all existing work is concerned with caching in deployed information retrieval systems, rather than for the purposes of aiding experimentation. 

In the remainder of this paper, we provide a background on PyTerrier (Section~2), followed by a description of our precomputation method for IR experiments (Section~3), explicit caching transformers (Section~4). Section~\ref{sec:exp} provides some demonstration experiments of the benefits of the precomputation and caching to an exemplar experiment; Section 6 discusses limitations and future work, while concluding remarks follow in Section~7.

\section{Information Retrieval Pipelines and Experiments in PyTerrier}
We provide an overview of the PyTerrier data model and operator language (Section~2.1), as well as the manner in which evaluation is conducted in PyTerrier (Section 2.2). Together these provide these necessary background for our understanding the functional aspects of our solutions described in Sections 3 and 4.

\subsection{Declarative Pipelines}

We now provide a short summary of the PyTerrier~\cite{DBLP:conf/ictir/MacdonaldT20} data model, the families of transformers, and operators for their combination. Firstly, let $Q(qid, query)$ be a relation type for a set of queries, and similarly $D(docno, text, ...)$ a relation type for a set of documents, \cm{with additional possible attributes}. From these principle  types, we can derive types such as for (i) documents ranked in response to a query $R(qid, docno, score, rank, ...)$; (ii) relevance assessments $RA(qid, docno, label)$. All relation types are extensible, in that extra columns can be added. In PyTerrier, these relation types can be instantiated as Pandas DataFrames, or as lists of dictionaries, both with the required and any optional attributes. Indeed, the exact choice of instantiation is left to the preferences of the developer of a particular class, and PyTerrier maps between the DataFrames and lists as needed. 

Transformations between these relation types are called {\em transformers}: for instance a transformer class fulfilling a retrieval role should expect data of type $Q$, and return type $R$. Typical families of pipeline stages can be expressed as mappings between these relations, for instance:
\begin{itemize}
\item {\em Retrieval}: $Q \to R$
\item {\em Reranking}: $R \to R$
\item {\em Query rewriting}: $Q \to Q$
\item {\em Document rewriting}: $D \to D$
\item {\em Pseudo-relevance feedback}: $R \to Q$
\item {\em Indexing}: $D \to \emptyset$ (a terminal operation)
\end{itemize}

Each transformer object, $t$, operates as a function, i.e. we can obtain a set of results for a given input by invoking $t$ on that input, $t(input)$.

As argued above, IR systems are more commonly phrased as pipelines. To combine different transformers, PyTerrier offers a number of operators defined on transformers, that allow them to be succinctly expressed. For instance $\gg$ is known as ``then'' or ``compose'', and is defined as:
\begin{align}
(t_1 \texttt{>>} t_2)(input) := t2(t1(input))
\end{align}
This allows many pipelines of transformers to be created, and easily understood. Indeed, rather than an imperative programming style, where a series of steps may be executed for a given set of data in sequence, here the entire pipeline is constructed before execution on a set of queries. The $\gg$ notation has been seen to be easily understandable and has been appearing as notation in many IR papers in recent years. Table~\ref{tab:operators} summarises all of the operators defined in PyTerrier. Each operator has relational algebra semantics, as explored in~\citet{DBLP:conf/ictir/MacdonaldT20}.

\begin{table}[bp]
    \centering
     \caption{PyTerrier operators for combining transformers.}    \label{tab:operators}
    \begin{tabular}{ccp{10cm}}
    \toprule
     Op.\ & Name & Description \\
    \midrule
        \texttt{>}\texttt{>} & \textit{then} & Pass the output from one transformer to the next transformer\\
        \texttt{\%} & \textit{rank cutoff} & Shorten a retrieved results list to the first $K$ elements \\
        \texttt{+} & \textit{linear combine} & Sum the query-document scores of the two retrieved results lists \\
        \texttt{*} & \textit{scalar product} & Multiply the query-document scores of a retrieved results list by a scalar \\
        \texttt{**} & \textit{feature union} & Combine two retrieved results lists as features \\
        \texttt{|} & \textit{set union} & Make the set union of documents from the two retrieved results lists \\
        \texttt{\&} & \textit{set intersection} & Make the set intersection of the two retrieved results lists \\

        \texttt{\string^} & \textit{concatenate} & Add the retrieved results list from one transformer to the bottom of the other \\
        \texttt{$\sim$} & \textit{cache} & Keep the results of this transformer on disk [now deprecated] \\
        
    \bottomrule
    \end{tabular}\vspace{-\baselineskip}
\end{table}

Of note from Table~\ref{tab:operators}, $\gg$ and $\%$ (rank cutoff) are the most commonly used operators. For example, we might want to apply rank cutoffs before different reranking stages:
\begin{lstlisting}
pipe = bm25 % 100 >> MonoT5() % 10 >> DuoT5()
\end{lstlisting}

Within Table~\ref{tab:operators}, $\sim$ is a caching operator, which is used to cache the results of a retrieving transformer to disk. This operates such that when a query is executed repeatedly, the results for that query can be retrieved directly from the cache, in order to speed up retrieval, particularly for experiments where a first-stage retrieval may be invoked repeatedly for the same query(ies). In brief, this is the only operator we have been unhappy with, and it is now deprecated for removal in a future release of PyTerrier. The semantics of how the caching operator is currently defined does not offer sufficiently fine-grained control - in particular, the type of caching that is most appropriate likely varies according to the family of transformer, which is not immediately obvious to the cache implementation. \cm{For example, a cache for a retriever should cache on the $qid$ attribute (the primary key of $Q$ datatype), and return the score for that query. However, caching on only the $qid$ would result in a rewritten query retrieving the same results as the original query, so $\langle qid, query \rangle$ would be a safer cache key. On the other hand, the cache for a cross-encoder reranker should cache on at least  $\langle qid, docno \rangle$, and, more safely $\langle qid, docno, query, text \rangle$, (in order to rescore when the query or text have been expanded).} More generically, the primary keys and the functional dependencies of the input and output types of a transformer are not exposed by a given transformer, and would be required for caching to operate properly. Such fine-grained control over caching cannot be achieved using only a unary caching operator to signify a cached transformer (indeed, the current caching operator assumes that all cached transformers are retrievers). For this reason, this paper discusses alternative caching strategies, \cm{addressing common experimental use cases easily, while also offering fine-grained control over caching behaviour.}

\subsection{Declarative Experiments}
PyTerrier also defines an experimentation abstraction for the purposes of evaluating different retrieval systems (including composed pipelines). In particular, the {\tt pt.Experiment()} function takes four key arguments: (i) a list of systems to be compared; (ii) the set of queries on which they should be evaluated (type $Q$); (iii) the set of relevance assessments to use to evaluate them (type $RA$); and (iv) the evaluation measures to compute. This invokes each system on the specified queries, and applies the specified evaluation measures on the outcome. An example experiment is shown below, which would evaluate the impact on the number of documents retrieved by BM25 on the nDCG@10 effectiveness of a pipeline involving MonoT5 and DuoT5 (recall that {\tt \%} denotes application of a rank cutoff):
\begin{lstlisting}
    pt.Experiment(
        [bm25 % k >> MonoT5() % 10 >> DuoT5() for k in [20, 50, 100, 200]],
        dataset.get_topics('test'),
        dataset.get_qrels('test'),
        [nDCG@10]
    )
\end{lstlisting}

The succinctness of the experimentation abstraction demonstrates its utility for researchers - indeed, we place considerable focus on developing memorable APIs that don't require researchers to regularly refer to documentation. Its also more succinct than an imperative workflow - its a single statement - no for loops, no different invocations for different stages of a ranking pipeline. Additional options for {\tt pt.Experiment()} allow the calculation of significance tests wrt.\ a baseline, application of multiple-testing correction (as recommended by~\citet{10.1145/3397271.3402426, 10.1145/3451964.3451976}), limiting batch sizes, etc. 

However, the example above also illustrates one of the challenges with a declarative workflow, in that the BM25 retriever would \cm{be} invoked for the topic set on each of the 4 pipelines. This is a considerable efficiency disadvantage compared to an imperative workflow, whereby a user may gather all BM25 results, before applying the cutoff. An alternative formulation is shown below - here BM25 is applied on the topics before ingestion into the experiment; an Identity transformer\footnote{The Identity transformer just returns its input, unchanged.} is used in place of BM25 to allow the results to be passed-through to a rank cutoff:
\begin{lstlisting}
    pt.Experiment(
        [pt.Transformer.identity() % k >> MonoT5() % 10 >> DuoT5() for k in [20, 50, 100, 200]],
        bm25(dataset.get_topics('test')), # <--- BM25 results as input, rather than queries 
        dataset.get_qrels('test'),
        [nDCG@10]
    )
\end{lstlisting}

\looseness -1 So while this is explicit, it somehow feels less appropriate - a reader of the code would not be naturally drawn to line 3 where \texttt{bm25} is invoked on the test queries; \cm{the clear separation between queries and systems has been lost}. The use of the Identity transformer also feels unnatural, and likely to confuse readers.

Instead, in the next section, we discuss how a different strategy whereby the same experiment is efficiently conducted, i.e. without repetitive invocations of BM25.

\begin{figure}
\centering
\includegraphics[width=0.6\linewidth]{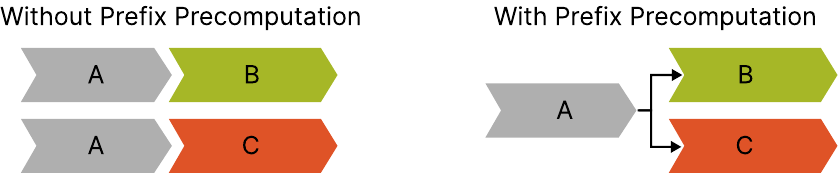}
\caption{A visual depiction of the prefix precomputation approach for two pipelines: A >> B and A >> C. With prefix precomputation, the common A prefix is identified and the results are used for the computation of the remainder of both pipelines, i.e. B \& C.}
\label{fig:prefix-precomp}
\end{figure}

\section{Prefix Precomputation in Comparative Experiments}\label{sec:common-prefix}
PyTerrier's operator-based language for expressing pipelines can be seen as closer to the conceptual design that one might write in a paper.\footnote{Indeed, we've noted an increasing number of papers using the $\gg$ notation to indicate composition of pipeline stages.} However, the logical implementation may differ. For example, consider a transformer $Retriever(index, k)$ that has a rank cutoff operation ($\%k'$) applied. A more efficient \cm{pipeline formulation} would be to apply the rank cutoff directly in the Retriever instance.\footnote{\cm{This is akin to a SQL query optimisation, where selection operations are moved earlier}.} PyTerrier supports a number of such optional {\em compile} operations, which allow apply a rewriting of the conceptual pipeline into a more efficient logical variant~\cite{DBLP:conf/ictir/MacdonaldT20} - \cm{i.e.~a syntactically different but semantically equivalent reformulation of a pipeline that executes more quickly}.

In this vein, let us consider an experiment comparing rerankers ($B$ and $C$) applied to the results of a retriever $A$. This would be instantiated as an experiment involving two pipelines $A \gg B$ and $A \gg C$. When performing a side-by-side evaluation of these pipelines, as mentioned in Section~2.2, there may be efficiency gains in pre-computing the results of $A$.

\looseness -1 To this end, PyTerrier now supports {\em prefix precomputation} when conducting experiments: here, any common prefix of all the evaluated pipelines is invoked once, and the results  applied on the remainder of the pipelines. This is shown visually in Figure~\ref{fig:prefix-precomp}. \cm{To expose this functionality} to the researcher, we simply add an optional argument to {\tt pt.Experiment()}, namely {\tt precompute\_prefix}, as shown below:
\begin{lstlisting}
    pt.Experiment(
        [bm25 % k >> MonoT5() % 10 >> DuoT5() for k in [20, 50, 100, 200]],
        dataset.get_topics('test'),
        dataset.get_qrels('test'),
        [nDCG@10],
        precompute_prefix=True # <---- enable precomputation 
    )
\end{lstlisting}

Our current implementation identifies the longest common prefix (LCP)\footnote{It has to be prefix, rather than the more general longest common subsequence, as pipelines are affected by their leftmost constituent transformation.} of a set of pipelines - efficient implementations of this algorithm can be instantiated that only assume that transformers have an equality property (i.e. we can test to see if two transformers are equal).

Formally, let $P$ be a set of transformer pipelines for a given experiment, where each pipeline $p_i \in P$ consists of different stages of a pipeline $t_{i,1} \gg \ldots \gg t_{i,||p_i||}$ where the number of stages in a pipeline is denoted by $||\cdot||$. Further, let $p[j]$ denote the $j$th stage of a pipeline, and $p[j..k]$ denote a range of transformer stages.  In applying the LCP algorithm, we identify a common prefix $\text{LCP}(P)$, as follows:
\begin{align}
\text{LCP}(P) = \arg\max_{cp} \left\{ ||cp|| \text{~s.t.~}  cp[j] == p_i[j] \text{  } \forall i, 1..j \right\}
\end{align}
Using $LCP(P)$, we can identify the {\em remainder} pipelines $\hat{P}$, where each constituent $\hat{p}_i$ is the remainder of the pipeline $p$ starting after the common prefix, i.e. $\hat{p}_i = p_i[||LCP(P)||.. ||p_i||]$. Then, for a given set of queries, $q$, evaluation naturally takes place by obtaining the results on the common prefix, i.e. $interim\_res = \text{LCP}(P)(q)$, followed by evaluation of each pipeline remainder, i.e. $\hat{p}_i(interim\_res)$.%

The astute reader may observe that there are possible experiments of \cm{$M>2$ pipelines}, where a common prefix is shared by only a subset, $2..M-1$. Our current implementation would not benefit the efficiency of such experiments. We postulate that this may be addressed by counting the coverage of all possible pipeline prefixes, but leave this further development to future work. 

That said, our experience with prefix precomputation thus far has been positive - \cm{the functionality works as expected, with no unexpected corner cases that required to be resolved. As a result}, we may we set this to be default setting for {\tt pt.Experiment()} in the future. Overall, we argue that the simplicity of prefix precomputation addresses a key efficiency disadvantage of the declarative workflow, while enabling end-to-end evaluation and retaining legibility of the experiments. It can also be seen as a marked progress between separation of a conceptual model of an IR experiment and its logical implementation. We are not aware of any previous work considering the programmatic decomposition of IR pipelines in this manner to benefit experimentation.

For more fine-grained control, in the next section we discuss a different type of caching, where the researcher wishes to retain full control over what is reused between pipelines or pipeline invocations.

\section{Explicit Caching Strategies}\label{sec:explicit}
\cm{We now} describe the explicit caching strategies provided by the pyterrier-caching\footnote{\url{https://github.com/terrierteam/pyterrier-caching}} package. Four strategies are provided to cover a variety of use cases: \cm{key-value caching (Section~4.1); caching for scorers/rerankers (Section~4.2); caching for retrievers (Section~4.3), and caching for indexing operations (Section~4.4).}

\subsection{\texttt{KeyValueCache}: Key-Value Caching}

\begin{figure}
\centering
\includegraphics{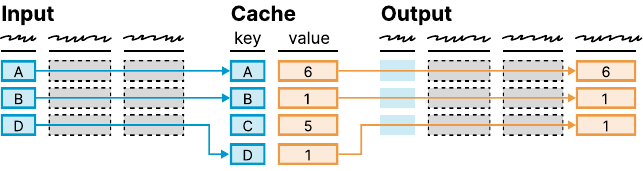}
\caption{The \texttt{KeyValueCache} maps each key (consisting of one or more input column) to a value (one or more output columns). It assumes rows are treated independently and that the values only depend on the keys.}
\label{fig:KeyValueCache}
\end{figure}

The \texttt{KeyValueCache} is a basic caching strategy that maps one or more ``key'' columns to one or more ``value'' columns. The cache operates on a row-by-row basis, working under the assumption that each row does not affect the results of other rows. A visual depiction of the cache is given in Figure~\ref{fig:KeyValueCache}. This formulation makes the \texttt{KeyValueCache} suitable for operations like Query Rewriting ($Q \to Q$) and Document Rewriting ($D \to D$). For example, a Doc2Query~\cite{DBLP:journals/corr/abs-1904-08375} transformer can be cached as follows:

\begin{lstlisting}
from pyterrier_caching import KeyValueCache
from pyterrier_doc2query import Doc2Query
dataset = pt.get_dataset('irds:msmarco-passage') # some dataset
model = Doc2Query(append=True)
cache = KeyValueCache('doc2query.cache', model, key='text', value='querygen')

# First index with Terrier: (fills cache with Doc2Query results)
index = pt.terrier.TerrierIndex('doc2query.terrier')
pipeline = cache >> index.indexer()
pipeline.index(dataset.get_corpus_iter())

# Indexing with PISA is now faster, since Doc2Query results are cached
from pyterrier_pisa import PisaIndex
index = PisaIndex('doc2query.pisa')
pipeline = cache >> index.indexer()
pipeline.index(dataset.get_corpus_iter())
\end{lstlisting}

\textbf{Implementation Details.} \texttt{KeyValueCache} is implemented as a SQLite database. Keys and values are encoded as blobs using Python's built-in \texttt{pickle} package. Rows with key misses in the database are passed along to the wrapped transformer to calculate their corresponding values, which are then inserted into the database.

\subsection{\texttt{ScorerCache}: Caching Scorer Results}

\looseness -1 Typical scorers (rerankers) operate by independently assigning a new relevance score for each document under the probability ranking principle~\cite{robertson1977probability}.\footnote{Note that this does not apply to some types of scorers, e.g., adaptive rerankers~\cite{DBLP:conf/cikm/MacAvaneyTM22} (which add new documents to the pool), or \cm{pairwise~\cite{DBLP:journals/corr/abs-2101-05667} and listwise}~\cite{DBLP:conf/emnlp/0001YMWRCYR23} rerankers (for which each score depends on the others present in the pool).} Given the prominence of this pattern and the need to re-assign the $rank$ column based on these new scores, the \texttt{ScorerCache} implements this special case of the general-purpose \texttt{KeyValueCache}. The $query$ and $docno$ columns serve as the key, and the $score$ column serves as the value, though this functionality can be overridden. An example demonstrating the value of the \texttt{ScorerCache} when caching the results of a MonoElectra~\cite{DBLP:conf/ecir/PradeepLZLYL22} model follows:

\begin{lstlisting}
from pyterrier_dr import ElectraScorer
from pyterrier_pisa import PisaIndex
dataset = pt.get_dataset('irds:msmarco-passage/dev/small')
index = PisaIndex.from_hf('macavaney/msmarco-passage.pisa')
scorer = dataset.text_loader() >> ElectraScorer()
cached_scorer = ScorerCache('electra.cache', scorer)

# Use the ScorerCache cache object just as you would a scorer
cached_pipeline = index.bm25() >> cached_scorer
cached_pipeline(dataset.get_topics())

cached_pipeline(dataset.get_topics()) # <-- all values are cached

# Will only compute scores for docnos that were not returned by bm25
another_cached_pipeline = index.qld() >> cached_scorer
another_cached_pipeline(dataset.get_topics())
\end{lstlisting}

\textbf{Implementation Details.} By default, \texttt{ScorerCache} uses the same SQLite strategy as \texttt{KeyValueCache}, while also re-assigning the $rank$ column based on the new scores. In some cases, a large proportion of a corpus is scored -- for instance, when exploring the effect of an exhaustive search strategy of cross-encoders~\cite{DBLP:journals/corr/abs-2306-09657}. In this case, a SQLite cache is inefficient because of the high repetition of document identifiers and other overheads. Therefore an alternative implementation (\texttt{DenseScorerCache}) is available. \texttt{DenseScorerCache} uses a HDF5 backend, with a separate npids\footnote{\url{https://github.com/seanmacavaney/npids}} file serving as a mapping between the docnos and their corresponding indexes in the storage array.

\subsection{\texttt{RetrieverCache}: Caching Retriever Results}

\begin{figure}
\centering
\includegraphics{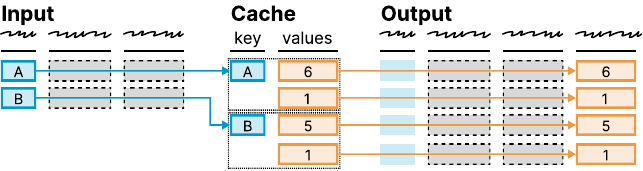}
\caption{The \texttt{RetrieverCache} maps each key (consisting of one or more input column) to a value (many rows over one or more output columns). It assumes input rows are treated independently and that the values only depend on the keys.}
\label{fig:RetrieverCache}
\end{figure}

Other operations --- most notably retrievers --- map each input row to multiple output rows. The \texttt{RetrieverCache} (shown visually in Figure~\ref{fig:RetrieverCache}) handles caching in this situation. The following code snippet shows how this cache can save on repetitive calls to a retriever:

\begin{lstlisting}
from pyterrier_caching import RetrieverCache
dataset = pt.get_dataset('irds:msmarco-passage/dev/small')
index = pt.terrier.TerrierIndex.from_hf('macavaney/msmarco-passage.terrier')
bm25_cache = RetrieverCache('path/to/cache', index.bm25())

bm25_cache(dataset.get_topics())
bm25_cache(dataset.get_topics()) # <-- all values are cached
\end{lstlisting}

\textbf{Implementation Details.} \texttt{RetrieverCache} is implemented using Python's \texttt{dbm} package. \sm{The \texttt{dbm} is a generic key-value database interface that is included in Python's standard library.} The keys to the DBM database are SHA256 hashed pickles of the keys, and the values are LZ4-compressed pickles of the value frames.

\subsection{\texttt{IndexerCache}: Caching Indexing Streams}

In some cases it is beneficial to store an entire sequence of inputs. This is especially true for indexing operations, such as document encoding with Learned Sparse Retrieval~\cite{DBLP:conf/ecir/NguyenMY23}. The following example shows how to cache SPLADE document representations~\cite{DBLP:conf/sigir/FormalPC21}. Note that unlike other caching operations, the \texttt{IndexerCache} acts as an indexer:

\begin{lstlisting}
from pyt_splade import Splade
from pyterrier_caching import IndexerCache
dataset = pt.get_dataset('irds:msmarco-passage')
splade = Splade()
cache = IndexerCache('splade.cache')
cache_pipeline = splade >> cache

# The following line will save the results of splade to splade.cache
cache_pipeline.index(dataset.get_corpus_iter())

# Now you can build multiple indexes over the results of splade without
# needing to re-run it each time
indexer1 = pt.terrier.TerrierIndex('splade.terrier').indexer()
indexer1.index(cache)
indexer2 = pyterrier_pisa.PisaIndex('./path/to/index.pisa').toks_indexer()
indexer2.index(cache)
\end{lstlisting}

\sm{Note that unlike the other caching components, \texttt{IndexerCache} captures a sequence of documents, where the order of records is potentially important~\cite{DBLP:conf/kdd/DhulipalaKKOPS16}. Consequently, it does not wrap a pipeline the way that other components do (i.e., \texttt{Splade() >> IndexerCache('path')} instead of \texttt{IndexerCache('path', Splade())}). The user may decide that the order is not important; in these cases, it} can also be used as a basic forward index, as it allows for efficient row lookups based on the $docno$ column (if present).

\textbf{Implementation Details.} \texttt{IndexerCache} stores the sequences as LZ4-compressed pickles of each input row. When  the cache object is iterated over, the sequence is decompressed row-by-row as a row generator. \sm{The index also captures the $docno$ column (if present), and stores it in an npids file to facilitate the forward index functionality.}

\subsection{Other Caching Features}

\looseness -1 Cache objects conform to PyTerrier's Artifact API, which allows them to be shared using HuggingFace, Zenodo, or other platforms. This can enable the sharing of computational resources across research groups.

\begin{lstlisting}
cache.to_hf('username/some-cache') # upload to HuggingFace
cache.to_zenodo() # upload to Zenodo
cache = pt.Artifact.from_hf('username/some-cache') # download from HuggingFace
cache = pt.Artifact.from_zenodo('1234') # download from Zenodo
\end{lstlisting}

All caches support a "temporary" cache mode, where a temporary cache directory is created and cleaned up when the object is deleted. This setting is applied by omitting the index path when creating the cache. We recommend using these as context managers so that the lifetime and cleanup of the temporary cache is well-defined. For example:

\begin{lstlisting}
from pyterrier_caching import RetrieverCache
dataset = pt.get_dataset('irds:msmarco-passage/dev/small')
index = pt.terrier.TerrierIndex.from_hf('macavaney/msmarco-passage.terrier')

# construct a temporary retriever cache
with RetrieverCache(retriever=index.bm25()) as bm25_cache:
    bm25_cache(dataset.get_topics())
    # second time faster due to caching
    bm25_cache(dataset.get_topics())
# (temporary cache deleted when the context manager exits)
\end{lstlisting}

All the caches also support a mode where the transformer is not provided. If there is a cache miss and the transformer is not provided, an exception is raised. Alternatively, the transformer can be constructed on-demand using the \texttt{Lazy} utility transformer, which only constructs the \cm{actual} transformer once if/when it is invoked. Both are helpful for situations where the user wants to avoid constructing the transformer object due to the resources that it would consume, e.g., a GPU for a neural scorer. For example:

\begin{lstlisting}
from pyterrier_caching import ScorerCache, Lazy
from pyterrier_dr import ElectraScorer

# scorer omitted (raises error on cache miss)
cached_scorer = ScorerCache('electra.cache')

# lazy scorer (only constructed once called)
lazy_scorer = Lazy(lambda: dataset.text_loader() >> ElectraScorer())
lazy_cached_scorer = ScorerCache('electra.cache', lazy_scorer)
\end{lstlisting}

\section{Demonstration Experiments}\label{sec:exp}
To demonstrate the benefit of the PyTerrier techniques described in this paper, we examine the response times of four different experimental settings. Our chosen experiment is based on the example pipelines discussed in Section 2.2, where we vary the number of documents retrieved by BM25 to determine the impact on MonoT5 and DuoT5. Such an experiment would use a list of pipelines expressed as follows: \texttt{[bm25 \% k >> MonoT5() \% 10 >> DuoT5() for k in [20, 50, 100, 200]]}.

The four different experimental settings are: (1) without using any caching, such that BM25 is executed four times on each query; (2) using precomputation (Section~3), such that BM25 is only executed once; (3) using a cold ScorerCache around MonoT5, such that MonoT5 is not executed more than once for a given query/document pair within the experiment; and (4) a re-run of the same setting where the ScorerCache is hot, such that MonoT5 reuses the results computed in (3). Finally, recall that DuoT5 is not amenable to caching in these pipelines, as the overall score of a document depends on the other retrieved documents for that query.

We execute these experiments on a machine with Intel Xeon Gold 5222 CPU @ 3.80GHz (16 cores) and an NVIDIA RTX 3090 GPU.  We use a Terrier backend for BM25, with the index stored on disk. We use two indices: MSMARCO v1 passage corpus using 43 queries from the TREC 2019 Deep Learning track; and MSMARCO V2 passage corpus using 53 queries from the TREC 2021  Deep Learning track. The Jupyter notebooks for executing these experiments can be found on the PyTerrier GitHub repo: \url{https://github.com/terrier-org/pyterrier/blob/master/examples/notebooks.md}.

The obtained timings are shown in Table~\ref{tab:experiments}, allowing the following observations: precomputation allows reducing the experimental execution time by 8\% on MSMARCO v1 and 28\% on MSMARCO v2 -- the small relative benefit on MSMARCO v1 is due to the small size of the MSMARCO v1 index and resulting fast BM25 retrieval (only 3 seconds for 43 queries); Caching of MonoT5 results shows a benefit of 27-41\%; Rerunning using a hot ScorerCache reduced overall time to 50-68\% - essentially executing BM25 once, and the reexecutions of DuoT5. Overall, the table supports the expected benefits of prefix precomputation and appropriate transformer caching, which are easily accessible through PyTerrier's \texttt{pt.Experiment} API and the the pyterrier-caching package. Prefix precomputation is more beneficial for more expensive shared prefixes of transformer pipelines in an experiment.

\begin{table}[]
    \centering
    \begin{tabular}{c|c|c|cc|cc}
         \toprule
          & & & \multicolumn{2}{c}{MSMARCO v1} & \multicolumn{2}{|c}{MSMARCO v2} \\
         \# & Precomputation & Cached MonoT5 & Execution Time & $\Delta$  & Execution Time & $\Delta$ \\
         \midrule
         (1) & \cross &  \cross & 3min 11s & - & 5min 48s & -\\
         (2) & \tick &  \cross & 2min 55s & 92\% & 4min 13s & 72\%\\
         (3) & \tick &  \tick~(cold) & 2min 19s & 73\% &3min 27s & 59\%\\
         (4) & \tick &  \tick~(hot) & 1min 36s & 50\% &2min 25s & 42\%\\
         \bottomrule
    \end{tabular}
    \caption{Execution times for an experiment comparing \texttt{[bm25 \% k >> MonoT5() \% 10 >> DuoT5() for k in [20, 50, 100, 200]]} on queries from the TREC 2019 \& 2021 Deep Learning tracks. Relative decreases in experiment execution time compared to row (1) are denoted by $\Delta$. }
    \label{tab:experiments}
\end{table}

\section{Limitations and Future Work}

\sm{We have described two approaches for caching in PyTerrier -- a precomputation approach that can be automatically applied to experiments, and explicit caching components that can be incorporated into indexing and retrieval pipelines. Although these additions provide a major improvement over the prior operator-based caching strategy, we see these as promising starting points, rather than final products, due to several shortcomings.}

\sm{Although the precomputation approach covers many practical experimental settings, it is not comprehensive. For instance, consider an ablation experiment where components are progressively added to a pipeline: (1) \texttt{A}, (2) \texttt{A >> B}, (3) \texttt{A >> B >> C}. The precomputation approach will only precompute \texttt{A} (since it's common to all three pipelines), even though also precomputing \texttt{A >> B} would be able to benefit both pipelines 2 and 3. Furthermore, the current strategy only supports precomputation across sequential pipelines; operations that occur as part of linear combinations, set operations, or feature combinations are not supported. We aim to assess which of these cases can be practically addressed and efficiently precomputed in future iterations of this feature -- potentially in combination with the \texttt{pipeline.compile()} functionality~\cite{DBLP:conf/ictir/MacdonaldT20}.}

\sm{The explicit caches also have limitations. Most notably, they rely on direct application by the researcher. This is by design, since current transformer implementations do not provide sufficient information to automatically infer the correct caching strategy. In the future, we may enhance the Transformer API to include this kind of information, e.g. the input and output columns, to ease the process of identifying the caching strategy to apply.\footnote{An added benefit is this information would also allow the automatic type-checking of pipelines.}}

\sm{Both precomputation and explicit caching make determinism assumptions; that is, the same input will yield the same output. This is not always the case, especially for components running on GPUs.\footnote{The non-determinism on GPUs arises from the non-deterministic order in which operations are assigned to the GPU, which can ultimately cause cascading differences in floating point operations.} In some sense, caching reduces the variability in experiments due to this noise. On the other hand, it means that such variations are not encountered through the course of experimentation, which could lead the researcher to infer a false sense of stability.}

\cm{Finally, precomputation is an first study of optimisation across multiple IR pipelines, and can be seen as similar to multi-query optimisation~\cite{Roy2009} in database management systems - we believe there may be other multi-query optimisation techniques from databases that can result in improved IR experimentation, which we leave to future work.}

\section{Conclusion}

We presented recent additions to the open-source PyTerrier platform to better facilitate caching. This involves two approaches: the precomputation of common pipeline prefixes when executing an experiment and new explicit result caching components. \cm{Demonstration experiments on MSMARCO v1 and v2 passage corpora concerning a pipeline with BM25, monoT5 and DuoT5 showed how precomputation and caching could reduce the execution time of a particular experiment by upto 41\% in a cold-cache setting.} We hope that these \cm{approaches} are intuitive for researchers to use, and will help reduce the computational cost of running experiments, promote GreenIR principles, and ease collaboration through shared caches.

\begin{acknowledgments}
We thank Jan Heinrich Merker for helpful feedback and suggestions on the pyterrier-caching package. We also thank Andrew Parry for help implementing the longest common prefix algorithm for prefix computation. \cm{Finally, we thank the anonymous reviewers for their detailed and thoughtful feedback.}

\end{acknowledgments}

\section*{Declaration on Generative AI}
During the preparation of this work, the authors used Grammarly for: Grammar and spelling check. After using these tools, the authors reviewed and edited the content as needed and take full responsibility for the publication’s content.

\bibliography{biblio}

\end{document}